# Shortest Path Discovery in the Multi-layered Social Network

Piotr Bródka, Paweł Stawiak, Przemysław Kazienko

Institute of Informatics, Wrocław University of Technology
Wyb.Wyspiańskiego 27, 50-370 Wrocław, Poland
piotr.brodka@pwr.wroc.pl, stawiak@gmail.com, kazienko@pwr.wroc.pl

*Abstract*— **Multi-layered social networks consist of the fixed set of nodes linked by multiple connections. These connections may be derived from different types of user activities logged in the IT system. To calculate any structural measures for multi-layered networks this multitude of relations should be coped with in the parameterized way. Two separate algorithms for evaluation of shortest paths in the multi-layered social network are proposed in the paper. The first one is based on pre-processing – aggregation of multiple links into single multi-layered edges, whereas in the second approach, many edges are processed 'on the fly' in the middle of path discovery. Experimental studies carried out on the DBLP database converted into the multi-layered social network are presented as well.**

## I. Introduction and Related Work

Complex networks are more and more important those days. With the growth of the Internet and various social networking systems the need to analysing them also is growing. There are many types of complex networked systems, e.g. physical systems (energy and transportation networks), virtual systems (Internet, WWW, telecommunication), biological networks, social networks, food webs and ecosystems [1]. In this paper we analyses social network, the concept of which, was first coined in 1954 by J. A. Barnes in [2]. The general concept of social network is simple and can be expressed as the finite set of actors (network nodes) and relationships (network edges) that link these actors. Although this concept appears to be quite obvious, almost every researcher describes it in a slightly different way and from different perspective [3], [4], [5], [6]. There are different domains of social networks: corporate partnership networks (law partnership) [7], scientist collaboration networks [8], movie actor networks, friendship network of students [9], company director networks [10], sexual contact networks [11], labour market [12], public health [13], psychology [14], etc.

The most interesting are social networks named online social networks [15], [16], web-based social networks [17.], computer-supported social networks [18] or virtual social networks because, due to an increase in technology we have continuous access to data from which those social network can be extracted, e.g.: bibliographic data [19], blogs [20], photos sharing systems like Flickr [21], e-mail systems [22], telecommunication data [23], [24], social services like Twitter [25] or Facebook [26], [27], video sharing systems like YouTube [28], Wikipedia [29] and many more. Researchers usually analyses only one kind of connections between users, while in most cases there is many different relationships. For example, Kazienko et al. investigated Flickr photo sharing system and have distinguished eleven types of relationships between users [21.]. A special type of social networks that allows to present many different activities is called a multi-layered social network [30]; it can be represented as a multi-graph [29], [31] Multi-layered social networks are much more harder to analyse than one-layered, because for them there is no well-known and widely-used social analysis methods and masseurs. In this paper, we focus on analysing the shortest paths Discovery in multi-layered social networks. The shortest paths problem is one of the most fundamental network optimization problems and algorithm for this problem have been studied for a long time [32], see e.g. [33] [34] [35] From all available algorithms in this article two of them was chosen as a base: Dijkstra for single shortest paths finding and Floyd-Warshall for all pairs shortest paths. SNA (social network analysis) provides measures in which we have to calculate first all pairs shortest paths in network. These are, for example beetweeness centrality or closeness centrality.

## II. Multi-layered Social Network

Definition 1: A multi-layered social network MSN is defined as a tuple $<V,E,L>$, where: $V$ – is a not-empty set of nodes (social entities: humans, groups of people); $E$ – is a set of tuples $<x,y,l>$, $x,y \in V$, $l \in L$, $x \neq y$ and for any two tuples $<x,y,l>$, $<x',y',l'> \in E$ if $x=x'$ and $y=y'$ then $l \neq l'$; $L$ – is a set of distinct layers.

Each tuple $<x,y,l>$ is an edge $e$ from $x$ to $y$ in layer $l$ in the multi-layered social network (MSN). The condition $x \neq y$ preserves from loops, i.e. reflexive connections from $x$ to $x$. There may exist only one edge $e=<x,y,l>$ from $x$ to $y$ in a given layer $l$. Note that any two nodes $x$ and $y$ may be connected with up to $|L|$ edges coming from different layers. All edges in MSN are directed and for that reason, $<x,y,l> \neq <y,x,l>$. Each layer corresponds to one type of relationship between users [30]. The examples of different relationship types can be friendship, family or work ties. The separate relationship types can also be defined based on distinct user activities towards some 'meeting objects', for example publishing photos, commenting photos, adding photos to favourites, etc, see [30], for details. Depending on variety of user actions logged in the database, we can have more or less layers in MSN.

The work was supported by: The Polish Ministry of Science and Higher Education, the development project, 2009-2011, The Polish Ministry of Science and Higher Education the research project, 2010-13 and Fellowship co-Financed by European Union within European Social Fund.

Nodes *V* and edges *e* from only one layer *l*∈*L* (such edges form set $E_l$) correspond to a simple, one-layered social network <*V*, $E_l$, {*l*}>.

A multi-layered social network MSN=<*V,E,L*> may be represented by a directed multi-graph. Hence, all the below proposed methods may also be applied to other kinds of complex networks that are characterized by means of multi-graphs.

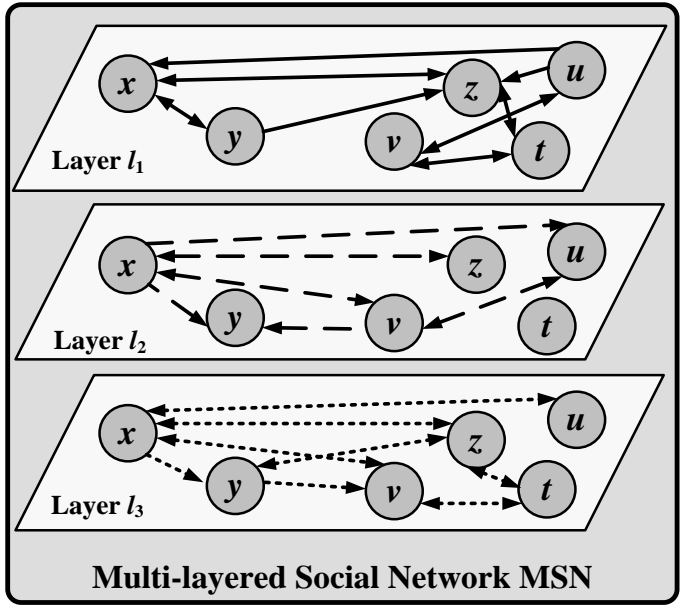

FIGURE 1 AN EXAMPLE OF THE MULTI-LAYERED SOCIAL NETWORK MSN

In Figure 1, the example of three-layered social network is presented. The set of nodes consists of {*x, y, u, v, z*} so there are five users in the network that can be connected with each other in three layers: $l_1$, $l_2$ and $l_3$. In layer $l_1$, eight relationships between users: <*x,y,$l_1$*>, <*y,x,$l_1$*>, <*x,z,$l_1$*>, <*z,x,$l_1$*>, <*y,z,$l_1$*>, <*u,z,$l_1$*>, <*u,v,$l_1$*>, <*v,u,$l_1$*> can be distinguished.

Neighbourhood $N^{out}(x)$ of a given node *x* for regular one-layered social network SN=<*V,E*> is defined as:

$$N^{out}(x) = \{y : <x, y> \in E\} \quad (1)$$

Multi-layered neighbourhood, *MN*(*x,α*), of a given node *x*∈*V* is a set of nodes, which are neighbours of node *x* on at least α layers in the MSN. $MN^{out}(x,α)$ is a multi-layered neighbourhood of node *x* with outgoing edges:

$$MN^{out}(x,\alpha) = \{y : card(\{l : <x, y, l> \in E\}) \geq \alpha\} \quad (2)$$

For MSN from Figure 1, we have $MN^{Out}(x,1)$={*u,v,y,z*}, $MN^{Out}(x,2)$={*u,v,y,z*}, $MN^{Out}(x,3)$={*y,z*}

### III. SHORTEST PATH DISCOVERY IN MULTI-LAYERED SOCIAL NETWORK

#### A. Transformation into Distance

To extract the shortest paths from MSN, some existing algorithms can be utilized. However, since there is no single edge between pairs of members in MSN, it is required to calculate the distance between any two neighbours in *MSN*. Basically, in most of the algorithms the shortest path is extracted based on the cost of transition from one node to another (negative connection), i.e. the greater cost the longer path value. On the other hand, in most social networks, the edges (relations) are considered to be positive, so, they express how close the two nodes are to each other, the greater weight the shorter path. That is why each typical 'positive weight', $w(x,y,l) \in [0,1]$ assigned to edges in MSN, has to be converted to 'negative' distance values by subtracting *w* from one, Figure 2. If there is no relationship between user *x* and user *y* then $w(x,y,l) = 0$.

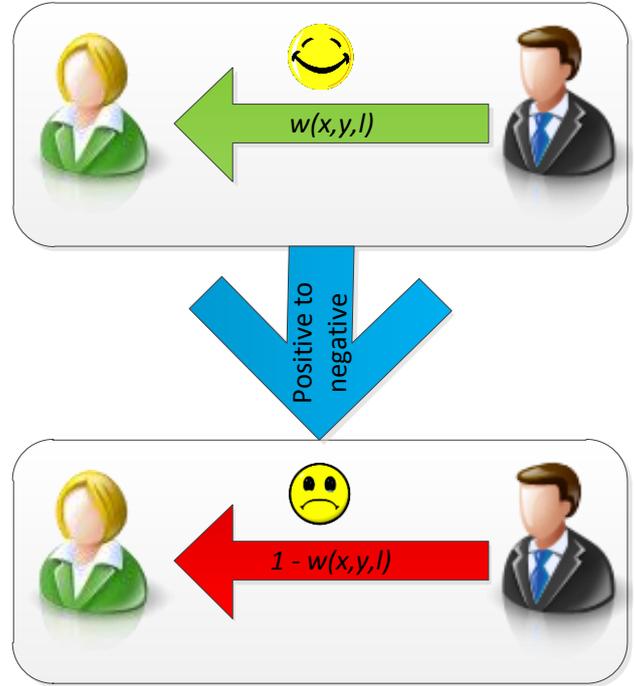

FIGURE 2 TRANSFORMATION OF 'POSITIVE' CLOSENESS SOCIAL RELATIONSHIP INTO 'NEGATIVE' DISTANCE (STRANGENESS).

The distance (strangeness) between *x* and *y* in MSN aggregated over all layers – *d*(*x,y*) is calculated using the weights $w(x,y,l) \in [0,1]$ of all existing edges <*x,y,l*> from *x* to *y*, i.e. from all layers. The sum of these weights is subtracted from the number of layers - |*L*| and normalized by the total number of layers in MSN:

$$d(x,y) = \frac{\sum_{l \in L}(1-w(x,y,l))}{|L|} = \frac{|L| - \sum_{l \in L}w(x,y,l)}{|L|} = $$
$$= 1 - \frac{\sum_{l \in L}w(x,y,l)}{|L|}. \quad (3)$$

Of course, if MSN has been built based on negative relations like: apathy, hatred, hostility, etc. the weight does not have to be subtracted from 1 to obtain the distance because it already reflects the distance.

B. *Multi-layered Edge*

Once, we have distance characterized, the multi-layered edge *ME* between two users can be defined. Overall, three different versions of multi-layered edge *ME* may be distinguished:

1. A multi-layered edge $ME^l$ based on the number of layers is computed in following way:

$$ME^l(x,y,\alpha) = <x,y>: |\{<x,y,l>: l \in L\}| \geq \alpha \quad (4)$$

A multi-layered edge $ME^l$ from *x* to *y* exists if there are at least $\alpha$ edges $<x,y,l>$ from *x* to *y* in MSN, i.e. $|\{<x,y,l>: l \in L\}| \geq \alpha$. The weight of such multi-layered edge is equal to the distance from *x* to *y*, $d(x,y)$, see Eq. 2.

2. A multi-layered edge $ME^d$ based on the distance between users is computed as follows:

$$ME^d(x,y,\beta) = <x,y>: d(x,y) \leq \beta \quad (5)$$

A multi-layered from *x* to *y* if the distance from *x* to *y* is lower than or equal $\beta$: $d(x,y) \leq \beta$. The weight of such an edge is equal to the distance from *x* to *y*, $d(x,y)$, see Eq. 2.

3. A multi-layered edge *ME* calculated based on the number of layers and on the distance between users is calculated in following way:

$$ME(x,y,\alpha,\beta) = <x,y>: |\{<x,y,l>: l \in L\}| \geq \alpha \wedge d(x,y) \leq \beta \quad (6)$$

A multi-layered edge *ME* from *x* to *y* if in the MSN exist at least $\alpha$ edges, $<x,y,l>$, from *x* to *y*, $|\{<x,y,l>: l \in L\}| \geq \alpha$ and simultaneously the distance from *x* to *y* is not greater than $\beta$, i.e. $d(x,y) \leq \beta$. The weight of such a multi-layered edge *ME* is equal to the distance from *x* to *y*, $d(x,y)$, see Eq. 2.

C. *Multi-layered Path*

The multi-layered path is a list of nodes and from each of nodes there is a multi-layered edge to the next node in the list. The length of such path from node *x* to *v* is a sum of distances of all multi-layered edges from the path.

The shortest multi-layered path is the path with the smallest length $SP(x,v)$ among all paths existing from node *x* to *v* in multi-layered social network.

D. *Shortest Path Calculation*

Now, the concept presented above can be applied to calculate shortest paths in two different approaches.

First approach is to calculate multi-layered edges and their length at the beginning of calculation. Next, any shortest path calculation algorithm can be utilized in the same way like for "normal" social network. For example, the "regular" Dijkstra algorithm provides all shortest paths with their lenths $SP(x,v)$ outgoing from a single node *x*. If applied to multi-layered edges, it is as follows:

---

**DAP – Dijkstra Algorithm with Preprocessing**

**Input**: $MSN=<V,E,L>$, $x \in V$ – source node
**Output**: the list of shortest paths outgoing from *x* and their lengths $SP(x,v)$ **for all** $v \in V$

Calculate distances $d(y,z)$ for each $y,z \in V$ using *Eq. 3*
Calculate all multi-layered edges *ME* using *Eq. 6*

$P=\phi$ (empty set)       /* P – set of nodes already processed */
$T=V$                      /* T – set of nodes to process */
$SP(x,v)=\infty$ for all $v \in V$
$SP(x,x)=0$
$pred(x,x)=0$              /* $pred(x,v)$ – predecessor of selected */
                           /* node *v* on the path from *x* to *v* */
**while** $P \neq V$ **do**
    **begin**
        $v=\text{argmin}\{SP(x,v) \mid v \in T\}$
        $P:=P \cup v$, $T:=T \setminus v$
        **if** $SP(x,v)=\infty$ **then** end while
        **for** $w \in N^{out}(v)$ **do**
            **if** $SP(x,w) > SP(x,v) + d(v,w)$ **then**
            **begin**
                $SP(x,w):= SP(x,v)+ d(v,w)$
                $pred(x,w)=v$
            **end**
    **end**

---

Second approach is to modify selected algorithm and process edges "on the fly". This approach is slightly better because additional multi-layered indicates can be added like multi-layered neighbourhood. Example on Dijkstra algorithm modified into multi-layered Dijkstra algorithm.

---

**MDA - Multi-layered Dijkstra Algorithm**

**Input**: $MSN=<V,E,L>$, $x \in V$ – source node
**Output**: the list of shortest paths outgoing from *x* and their lengths $SP(x,v)$ **for all** $v \in V$

$P=\phi$ (empty set)
$T=V$
$SP(x,v)=\infty$ for all $v \in V$
$SP(x,x)=0$
$pred(x,x)=0$
**while** $P \neq V$ **do**
    **begin**
        $v=\text{argmin}\{SP(x,v) \mid v \in T\}$
        $P:=P \cup v$,
        $T:=T \setminus v$
        **if** $SP(x,v)=\infty$ **then** end while
        **for** $w \in MN^{out}(v,\alpha)$ **do**
            **if** $SP(x,w) > SP(x,v) + d(v,w)$ **then**
            **begin**
                $SP(x,w):= SP(x,v)+ d(v,w)$

```
            pred(x,w)=v
        end
    end
```

## IV. Experiments

### A. Multi-layered Social Network Preparation

The DBLP Computer Science Bibliography dataset (http://www.informatik.uni-trier.de/~ley/db/) was used for experimental study. The data set was split for two parts to avoid performance problems. Due to computational complexity of shortest paths discovery ($O(n^3)$), where *n* is the total number of nodes, the file was split in The zipped file with the data is available at: http://www.zsi.pwr.wroc.pl/~kazienko/DBLP/asonam2011_dblp.zip.

Both one-layered and multi-layered social Network was created using the multi-layered social network creation process described in [30] and [36]. Additionally in order to prepare MSN for shortest path calculation the relationship strength calculated based on positive relations was converted into negative relationship (See section Shortest Path Discovery in Multi-layered Social Network).

### B. One-layered Social Network Analysis

At the beginning the simple one-layered social network was extracted. It consists of: one layer called cited publication author - cited publication author, 6 020 users and 348 206 relations. Next, for one-layered social network, the "regular" Dijkstra algorithm and the multi-layered Dijkstra $ME(x,y,\alpha,\beta)$ was calculated for $\alpha=1$ and $\beta=0.0$ was calculated.

Both "regular" Dijkstra algorithm and $ME(x,y,\alpha,\beta)$ has calculated 5,925 shortest paths from the node with id=860817. Moreover all shortest paths were equals.

Furthermore both "regular" Floyd Warshall algorithm and multi-layered Floyd Warshall algorithm $ME(x,y,\alpha,\beta)$ for $\alpha=1$ and $\beta=0.0$ was calculated and both have returned 35,111,837 equal shortest paths. Additionaly the computation time was almost the same. For small networks $ME(x,y,\alpha,\beta)$ computation time was up to 3% greater than "regular" algorithms and for big one there was no difference. The results returned by all algorithms was equal was equal, so is possible to use $ME(x,y,\alpha,\beta)$ in both one-layered and multi-layered network. Below is sample result from comparison.

### C. Multi-layered Social Network Analysis

After one-layered network analysis the multi-layered social network was created: It consist from 424,455 nodes and three layers presented in Table 1.

The multi-layered Dijkstra for $ME(x,y,\alpha,\beta)$ was calculated for $\alpha = \{1, 2, 3\}$ and $\beta = \{1, 0.975, 0.875, 0.667, 0.5, 0.333\}$. The results are presented in Table 2.

TABLE 1. LAYERS AND THE NUMBER OF EDGES ON EACH LAYER

| Layer index | Layer - relationships from different kinds of activities | no. of relations |
|---|---|---|
| 1 | publication author - publication author | 2,582,744 |
| 2 | publication author - cited publication author | 2,963,092 |
| 3 | cited publication author - cited publication author | 348,206 |

TABLE 2. THE RESULTS OF MULRI-LAYERED DIJKSTRA CALCULATION FOR $ME(x,y,\alpha,\beta)$

| Start Node | α | β | No. of routes | Avg. path length | Min path length | Max path length | Avg. no. of handshakes | No. of neighbourhoods | Percent of connected nodes |
|---|---|---|---|---|---|---|---|---|---|
| 238369 | 1 | 1 | 361,440 | 4.819 | 0.640 | 21.218 | 7.719 | 19 | 85% |
| 238369 | 2 | 1 | 361,282 | 5.113 | 0.640 | 21.787 | 7.750 | 19 | 85% |
| 238369 | 3 | 1 | 3,811 | 8.108 | 0.888 | 24.435 | 8.350 | 9 | 68% |
| 238369 | 1 | 0.975 | 287,535 | 28.267 | 0.640 | 70.386 | 44.083 | 19 | 68% |
| 238369 | 1 | <0.875 | 0 | 0 | 0 | 0 | 0 | 0 | 0% |
| 469619 | 1 | 1 | 361,440 | 4.609 | 0.306 | 21.165 | 8.133 | 12 | 85% |
| 469619 | 2 | 1 | 361,282 | 7.845 | 0.606 | 27.332 | 12.346 | 10 | 85% |
| 469619 | 3 | 1 | 0 | 0 | 0 | 0 | 0 | 0 | 0% |
| 469619 | 1 | 0.975 | 287,535 | 48.235 | 0.306 | 93.534 | 77.984 | 12 | 68% |
| 469619 | 1 | <0.875 | 0 | 0 | 0 | 0 | 0 | 0 | 0% |
| 135846 | 1 | 1 | 361,440 | 3.800 | 0.333 | 19.629 | 6.438 | 806 | 85% |
| 135846 | 2 | 1 | 361,282 | 4.153 | 0.656 | 22.520 | 6.427 | 690 | 85% |
| 135846 | 3 | 1 | 3,811 | 6.354 | 0.952 | 23.518 | 6.484 | 59 | 68% |
| 135846 | 1 | 0.975 | 287,534 | 40.998 | 0.952 | 95.953 | 62.965 | 3 | 68% |
| 135846 | 1 | <0.875 | 0 | 0 | 0 | 0 | 0 | 0 | 0% |
| 126314 | 1 | 1 | 361,440 | 4.711 | 0.314 | 20.914 | 7.782 | 16 | 85% |
| 126314 | 2 | 1 | 361,282 | 4.944 | 0.604 | 21.484 | 7.744 | 14 | 85% |
| 126314 | 3 | 1 | 3,811 | 8.405 | 0.604 | 20.018 | 8.987 | 1 | 68% |
| 126314 | 1 | 0.975 | 287,535 | 51.099 | 0.604 | 120.776 | 79.182 | 14 | 68% |
| 126314 | 1 | 0.875 | 1 | 0.604 | 0.604 | 0.604 | 1 | 1 | 0% |
| 126314 | 1 | 0.667 | 1 | 0.604 | 0.604 | 0.604 | 1 | 1 | 0% |
| 126314 | 1 | <0.5 | 0 | 0 | 0 | 0 | 0 | 0 | 0% |
| 286653 | 1 | 1 | 361,440 | 4.581 | 0.622 | 20.907 | 7.641 | 25 | 85% |
| 286653 | 2 | 1 | 361,282 | 4.942 | 0.622 | 22.140 | 7.708 | 25 | 85% |
| 286653 | 3 | 1 | 5 | 1.317 | 0.768 | 1.708 | 1.6 | 2 | 0% |
| 286653 | 1 | 0.975 | 287,534 | 40.644 | 0.622 | 95.600 | 62.963 | 3 | 68% |
| 286653 | 1 | <0.875 | 2 | 0.768 | 0.768 | 0.768 | 1 | 2 | 0% |
| 307208 | 1 | 1 | 361,440 | 4.366 | 0.331 | 20.280 | 7.180 | 55 | 85% |
| 307208 | 2 | 1 | 361,282 | 4.641 | 0.615 | 21.511 | 7.235 | 53 | 85% |
| 307208 | 3 | 1 | 3,811 | 9.896 | 0.880 | 23.400 | 10.221 | 5 | 68% |
| 307208 | 1 | 0.975 | 287,534 | 39.598 | 0.615 | 94.637 | 61.841 | 9 | 68% |
| 307208 | 1 | <0.875 | 0 | 0 | 0 | 0 | 0 | 0 | 0% |
| 32525 | 1 | 1 | 361,440 | 4.718 | 0.262 | 20.691 | 8.106 | 2 | 85% |
| 32525 | 2 | 1 | 361,282 | 4.996 | 0.262 | 21.265 | 8.149 | 2 | 85% |
| 32525 | 3 | 1 | 1 | 0.262 | 0.262 | 0.262 | 1 | 1 | 0% |
| 32525 | 1 | 0.975 | 287,534 | 32.729 | 0.262 | 73.272 | 51.549 | 2 | 68% |
| 32525 | 1 | 0.875 | 3 | 0.522 | 0.262 | 0.900 | 1.333 | 2 | 0% |
| 32525 | 1 | 0.667 | 1 | 0.262 | 0.262 | 0.262 | 1 | 1 | 0% |
| 32525 | 1 | 0.5 | 1 | 0.262 | 0.262 | 0.262 | 1 | 1 | 0% |
| 32525 | 1 | 0.333 | 1 | 0.262 | 0.262 | 0.262 | 1 | 1 | 13% |
| 84388 | 1 | 1 | 361,440 | 4.596 | 0.286 | 20.576 | 8.083 | 4 | 85% |
| 84388 | 2 | 1 | 361,282 | 5.023 | 0.310 | 22.713 | 8.014 | 3 | 85% |
| 84388 | 3 | 1 | 3,811 | 9.316 | 0.310 | 26.662 | 10.300 | 1 | 68% |
| 84388 | 1 | 0.975 | 287,534 | 29.442 | 0.286 | 69.983 | 47.554 | 4 | 68% |
| 84388 | 1 | 0.875 | 3 | 0.460 | 0.310 | 0.536 | 1 | 3 | 0% |
| 84388 | 1 | 0.667 | 1 | 0.310 | 0.310 | 0.310 | 1 | 1 | 0% |
| 84388 | 1 | 0.5 | 1 | 0.310 | 0.310 | 0.310 | 1 | 1 | 0% |
| 84388 | 1 | 0.333 | 1 | 0.310 | 0.310 | 0.310 | 1 | 1 | 13% |

The results presented in Table 2 and Figure 3 shows that both parameters α and β have different properties. Parameter α is better when the number of relations is more important than their strength. On the other hand β is based on the strength of relationship but indirectly also on their number (because it is average).

The multi-layered Dijkstra for $ME(x,y,\alpha,\beta)$, $\alpha=1, \beta=1.0$ and node with id = 238369 returned 361,440 shortest paths. The same algorithm for $\alpha=2, \beta=1.0$ calculated 361,282 shortest paths and for $\alpha=3, \beta=1.0$ displayed only 3,811 shortest paths for these same node. So with increase of α some paths changes or disappears. For example, there exist a path from node 238369 to node 423666 across node 385483 for $\alpha=1, \beta=1.0$ when for $\alpha=2, \beta=1.0$ there is a path only from node 238369 to node 385483 and connection between 385483 and 423666 disappears. For $\alpha=3, \beta=1.0$ even connection between 238369 and 385483 vanish.

Another interesting case is that for the node 343851 and $\alpha=1, \beta=1.0$ the path to node 32525 was across nodes 281873, **203749, 109770, 108044**, 276264 and its length was 3.35 while for $\alpha=2, \beta=1.0$ the path was across 281873, **241190, 204096, 169332**, 276264 and its length was 3.66.

In case of β there is more linear decrease of nodes in social network than in α case. But decreasing β very fast creates a sparse social network. For example, calculation with $\alpha=3, \beta=1.0$ has average no. of neighbourhood about 4 and for $\alpha=1, \beta=0.5$ it is just 1. Moreover in most cases for β<0.667 there was so few multi-layered edges that it is impossible to find more than few shortest paths. It may indicate that the network consist of many links with low relationship strength.

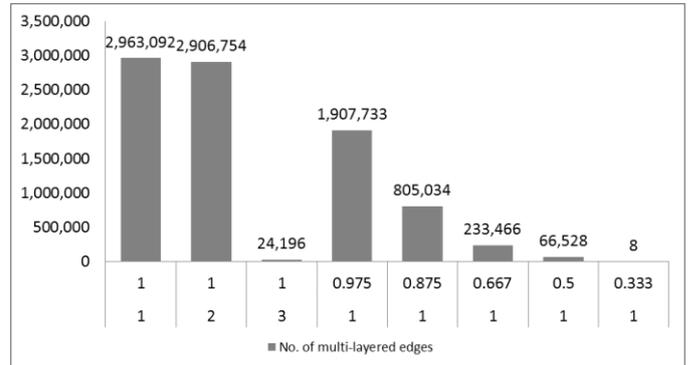

FIGURE 3 NOUMBER OF MULTILAYERED EDGES $ME^{\alpha\beta}(x,y,\alpha,\beta)$ FOR DIFFERENT VALUES OF $\alpha$ (on the lower x Axis) AND $\beta$ (on the upper x Axis)

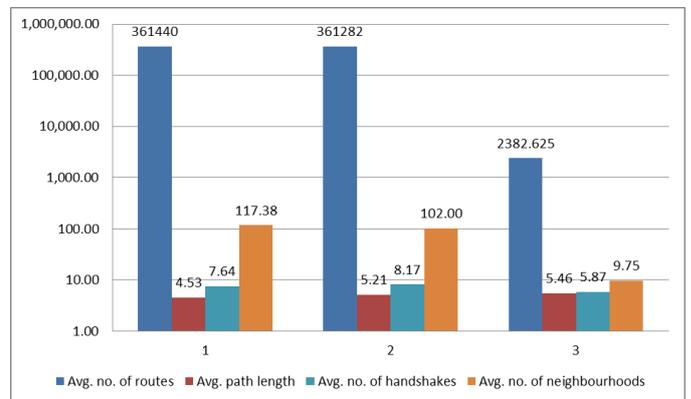

FIGURE 4 BASIC STATISTIC FOR $\alpha$ (x Axis)

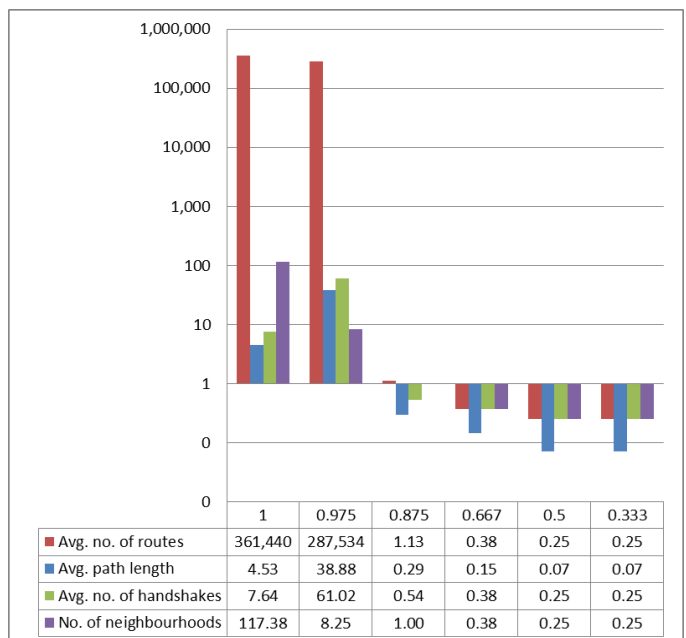

FIGURE 5 BASIC STATISTIC FOR $\beta$ (x Axis)

## V. CONCLUSIONS AND FUTURE WORK

To be able to process a multi-layered social network, we need to solve the problem of multiple connections between pairs of nodes. One of the possible approaches is to aggregate many links into single multi-layered edges *ME*. Three basic methods of such aggregation are presented in the paper: $ME^l$ based on the number of layers, $ME^d$ based on distance and combined multi-layered edges *ME*. The aggregations have incorporated two thresholds $\alpha$ and $\beta$, which enable to filter weak multiple connections, i.e. with the small number of links ($\alpha$ in $ME^l$) or with the small average weight of links ($\beta$ in $ME^d$) or both (in *ME*).

Based on the aggregated network (with multi-layered edges *ME*) some regular algorithms may be applied. For that purpose a Dijkstra algorithm with pre-processing DAP was proposed for shortest paths extraction. In another approach, the multi-layered social network is analysed without any prior processing: multiple edges are processed while shortest paths are computed. A multi-layered Dijkstra algorithm MDA was developed as an instance of algorithms with incorporated multiple connections processing.

Based on experiments carried out on the real multi-layered social network extracted from the DBLP database, we can find out that parameters $\alpha$ and $\beta$ strongly influence both quantity of paths as well as their average lengths.

Discovery of shortest paths is just introduction to vast number of multi-layered social network analysis measures like multi-layered beetweeness centrality or multi layered closeness centrality. Having shortest paths in multi-layered social network extracted, we can calculate many other measures so the future work will focus on adaptation of beetweeness and closeness centrality measures to multi-layered social network.